\documentclass[11pt]{article}
\newcommand{\be}{\begin{equation}}
\newcommand{\ee}{\end{equation}}
\newcommand{\ba}{\begin{array}}
\newcommand{\ea}{\end{array}}

\newcommand{\bea}{\begin{eqnarray*}}
\newcommand{\eea}{\end{eqnarray*}}
\newcommand{\bean}{\begin{eqnarray}}
\newcommand{\eean}{\end{eqnarray}}
\newcommand{\proof}{\vspace{1ex}\noindent{\em Proof}. \ }
\def\ds{\displaystyle}

\newcommand{\R}{\mbox{\bf R}}

\newtheorem{lem}{Lemma}[section]

\newtheorem{thm}{Theorem}[section]
\newtheorem{prop}{Proposition}[section]
\def\Box{\leavevmode\vbox{\hrule
     \hbox{\vrule\kern5pt\vbox{\kern5pt}%
           \vrule}\hrule}}
\newcommand{\square}{\hfill$\Box$}

\begin{document}
\title{On the perturbation of the electromagnetic energy due to the presence
 of inhomogeneities with small diameters}

\author{Christian DAVEAU   \and Abdessatar KHELIFI }
\date{}
\maketitle \abstract{ We consider solutions to the time-harmonic
Maxwell problem in $\R^3$. For such solution we provide a rigorous
derivation of the asymptotic expansions in the practically
interesting situation, where a finite number of inhomogeneities of
small diameter are imbedded in the entire space. Then, we describe
the behavior of the electromagnetic energy caused by the presence
of these inhomogeneities. }

\begin{center}
{\bf Sur la perturbation de l'\'energie \'electromagn\'etique due
\`a la pr\'esence des inhomog\'en\'eit\'es de petits diam\`etres }
\end{center}

\begin{center}
{\bf R\'esum\'e}
\end{center}
Nous consid\'erons des solutions du probl\`eme harmonique de
Maxwell dans $\R^3$. pour une telle solution nous obtenons des
formules asymptotiques rigoureuses  qui sont dues \`a la
pr\'esence d'un nombre fini d'inhomog\'en\'eit\'es avec petit
diam\`etre dans l'espace entier. Puis, nous d\'ecrivons le
comportement de l'\'energie \'electromagn\'etique provoqu\'e par
la pr\'esence de ces inhomog\'en\'eit\'es.

\section*{Version fran\c caise abr\'eg\'ee} Dans cette Note, on suppose que dans
 $\R^3$ on a $m$ inhomog\'en\'eit\'es $\{z_j +
\alpha B_j\}_{j=1}^m$, o\`u $\alpha$ est un petit param\`etre,
$B_j \subset \R^3$ est un ouvert born\'e et les points
$\{z_j\}_{j=1}^m$ v\'erifient l'hypoth\`ese (\ref{f1}). Soient la
perm\'eabilit\'e magn\'etique $\mu_\alpha$ et la permittivit\'e
\'electrique $\varepsilon_\alpha$ de forme (\ref{murhodef}).\\

L'objectif du travail d\'ecrit dans cette Note est de voir comment
les solutions perturb\'ees du probleme~(\ref{maxwell1}) se
comportent quand un nombre fini d'inhomog\'en\'eit\'es $\{z_j +
\alpha B_j\}$ de petits diam\`etres sont introduites dans l'espace
entier $ \R^3$. Ceci nous am\`ene, \'egalement, \`a \'etudier
l'\'evolution de l'\'energie \'electromagn\'etique correspondante
selon cette d\'eformation du milieu de propagation.\\ Nous
commençons notre analyse dans la section $2$ en d\'erivant
rigoureusement les d\'eveloppements asymptotiques des champs
\'electrique et magn\'etique ce qui sont
  {\it uniform\'ement valid\'es dans l'espace}. Ces formules asymptotiques sont
   construites par {\it la m\'ethode de d\'eveloppements asymptotiques assorties}.
    Concernant cette
 m\'ethode, le lecteur peut consulter \cite{cole,il}. En domaine
 born\'e, on peut voir aussi les travaux \cite{ammari-iak-less,
 ammari-khelifi, ammari-moskow-vogelius, ammari-darko}. Le terme d'ordre
 principale, dans des d\'eveoppements asymptotiques analogues (mais pour le cas
d'un domaine born\'e), a \'et\'e d\'eriv\'e par Vogelius et Volkov
\cite{VV} et Ammari et al. \cite{ammari-vog-darko}. Nos formules
asymptotiques utilisent des tenseurs de polarisation associ\'es
\`a des inhomog\'en\'eit\'es \'electromagn\'etiques qui semblent
\^etre des g\'en\'eralisations normales des tenseurs qui ont
\'et\'e pr\'esent\'es par Schiffer et Szeg\"{o} \cite{SS} et
compl\`etement \'etudi\'e par beaucoup
d'autres auteurs \cite{ammari-kang, CMV, FV,MS}.\\
Le plan de cette Note est comme suit: \\ Dans la section $1$, Nous
formulons le probl\`eme principale dans ce travail. Dans la
section $2$, supposant que les champs \'electromagn\'etiques
$({\bf E}_{\epsilon}, {\bf H}_{\epsilon}) $ sont dans des espaces
de Sobolev convenables, nous obtenons \`a travers la formule de
repr\'esentation int\'egrale de Lippman-Schwinger des
d\'eveloppements asymptotiques uniformes pour les champs
\'electromagn\'etiques et la densit\'e d'\'energie. Puis, nous
formulons une estimation pour l'\'energie \'electromagn\'etique ce
qui nous apporte aux r\'esultats de stabilit\'e n\'eanmoins la
pr\'esence de petites inhomog\'en\'eit\'es.

\section{Problem formulation}
We suppose that there is a finite number of electromagnetic
inclusions in $\R^3$, each of the form $z_j+\alpha B_j$, where
$B_j\subset\R^3$ is a bounded and $\mathcal{C}^{\infty}$-domain
containing the origin. The total collection of inhomogeneities is:
$$
\mathcal{B}_{\alpha}=\ds\cup_{j=1}^{m}\{z_j+\alpha B_j\}.
$$
 The points $z_j \in \R^3, j=1, \ldots, m,$ which determine the
  location of the inhomogeneities, are assumed to satisfy the
  following inequality:
\be \label{f1}
 | z_j  - z_l | \geq c_0 > 0, \forall \; j \neq l.
\ee Let $\mu_0$ and $\varepsilon_0$ denote the permeability and
the permittivity of the free space, $\omega>0$ is a given
frequency and $k=\omega\sqrt{\varepsilon_0\mu_0}>0$ is the wave
number. We denote by $x=(x_1,x_2,x_3)$ the cartesian coordinates
in $\R^3$. We shall assume that {\it $\mu_0>0$ and
$\varepsilon_0>0$ are positive constants}. Let $\mu_j>0$ and
$\varepsilon_j>0$ denote the permeability and the complex
permittivity of the j-th inhomogeneity, $z_j+\alpha B_j$; these
are also assumed to be positive constants. Introduce the
piecewise-constant magnetic permeability
\begin{equation}
\mu_\alpha(x):=\left \{ \begin{array}{*{2}{l}}
 \mu_0,\;\;& x \in \R^3 \setminus \bar {\cal B}_\alpha,  \\
 \mu_j,\;\;& x \in z_j+\alpha B_j, \;j=1 \ldots m.
\end{array}
\right . \label{murhodef}
\end{equation}
The piecewise-constant electric permittivity
$\varepsilon_\alpha(x)$ is defined analogously. If we allow the
degenerate case $\alpha =0$, then the function $\mu_0(x)$ (resp.
$\varepsilon_0(x)$) equals the constant $\mu_0$
(resp. $\varepsilon_0$).\\
We suppose now that the collection of inclusions
$\mathcal{B}_{\alpha}$ is within an open and fictive subset
$\Omega\subset\R^3$ and the source current density ${\bf J}_{s}$
is at position in $\R^3 \setminus \bar {\Omega}$.\\

Let $({\bf E}_{\alpha},{\bf H}_{\alpha})\in\R^3\times\R^3$ denote
the time-harmonic electromagnetic fields in the presence of the
electromagnetic inclusions ${\cal B}_\alpha$. These time-harmonic
fields \cite{hazard-lenoir,nedelec} are the solutions of the
Maxwell's equations
\begin{equation}
\left \{ \begin{array}{*{4}{l}}
 \nabla\times {\bf
E}_{\alpha}= i\omega\mu_{\alpha}{\bf
H}_{\alpha},\;\;& \mbox{  in }\R^3,  \\
\nabla\times {\bf H}_{\alpha}=-i\omega\varepsilon_{\alpha}{\bf
E}_{\alpha}+{\bf J}_{s},\;\;& \mbox{  in }\R^3,  \\
\nu\times {\bf E}_{\alpha} \quad \mbox{  and } \nu\times {\bf
H}_{\alpha}\;\;& \mbox{ are continuous
across }\partial(z_j+\alpha B_j),  \\
\ds\lim_{|x|\to \infty}|x|\big[ \nabla\times {\bf
E}_{\alpha}-ik\frac{x}{|x|}\times {\bf E}_{\alpha} \big]=0 \;\;&
\mbox{and }\ds\lim_{|x|\to \infty}|x|\big[ \nabla\times {\bf
H}_{\alpha}-ik\frac{x}{|x|}\times {\bf H}_{\alpha} \big]=0 .
\end{array}
\right . \label{maxwell1}
\end{equation}
Here $\nu$ denotes the outward unit normal to $\partial(z_j+\alpha
B_j)$. We eliminate the magnetic field from the above equations by
dividing the first equation in~(\ref{maxwell1}) by $\mu_{\alpha}$
and taking the curl to obtain the following system of equations
for ${\bf E}_{\alpha}$:
\begin{equation}
\left \{ \begin{array}{*{3}{l}}
 \nabla\times \mu_{\alpha}^{-1}\nabla\times{\bf
E}_{\alpha}-\omega^2\varepsilon_{\alpha}{\bf
E}_{\alpha}=i\omega{\bf J}_{s},\;\;& \mbox{  in }\R^3,  \\
\nu\times {\bf E}_{\alpha}\;\;& \mbox{ is continuous
across }\partial(z_j+\alpha B_j),  \\
\ds\lim_{|x|\to \infty}|x|\big[ \nabla\times {\bf
E}_{\alpha}-ik\frac{x}{|x|}\times {\bf E}_{\alpha} \big]=0.&
\end{array}
\right . \label{maxwell2}
\end{equation}
Having found the electric field ${\bf E}_{\alpha}$, we then obtain
the magnetic field ${\bf H}_{\alpha}$ through the formula
$${\bf H}_{\alpha}= \frac{1}{i \omega \mu_{\alpha}}
 \nabla\times { {\bf E}_{\alpha}}. $$
The electromagnetic energy is defined by:
 \begin{equation}\label{energy-def}
 \mathcal{E}_\alpha(t):=\frac{1}{2}\ds\int_{\R^3}
 (\varepsilon_\alpha|{\bf E}_{\alpha}(x,t)|^{2}+
 (\mu_\alpha)^{-1} |{\bf H}_{\alpha}(x,t)|^{2})dx,\quad \mbox{ for }t\geq 0,
 \end{equation}
 and it is not hard to see that $\mathcal{E}\in\mathcal{C}^1(\R^+)$.\\
 The finiteness of the electromagnetic energy $\mathcal{E}_\alpha$ requires that both
the electric and the magnetic field belongs to a space of fields
with square integrable {\bf curls}:
\[
 H({\bf curl};\R^3):=
 \{{\bf a}\in L^2(\R^3)^3 ;\quad {\bf curl}~{\bf a}\in L^2(\R^3)^3
 \}.
\]


It can be shown that there exists a unique solution $ {\bf
E}_{\alpha}\in
\mathcal{C}^1(\R^+;L^2(\R^3))\cap(\mathcal{C}^0(\R^+;
 H({\bf curl};\R^3)))$ to the problem (\ref{maxwell2}), and this
solution satisfies the following Lippman-Schwinger integral
representation formula .

\begin{lem}\label{lip} Let ${\bf E}_{\alpha}$ be the solution
 of the problem (\ref{maxwell2}). Then, the following integral representation
 formula holds
\begin{equation}\label{schinger1}
{\bf E}_{\alpha}(x)={\bf
E}_{0}(x)+\ds\sum_{j=1}^{m}\ds\int_{z_j+\alpha
B_j}\big(-i\omega(\mu_j -\mu_0)\nabla^{\prime}\times{\bf
\mathcal{G}}(x,x^{\prime})\cdot{\bf H}_{\alpha}(x^{\prime})
\end{equation}
$$+\omega^2\mu_0(\varepsilon_j -\varepsilon_0){\bf
\mathcal{G}}(x,x^{\prime})\cdot{\bf
E}_{\alpha}(x^{\prime})\big)~dx^{\prime}.$$
\end{lem}
The $3\times 3$ matrix valued function ${\bf \mathcal{G}}$ means
the Green's function solution to
$$
\left \{ \begin{array}{*{2}{l}}
 \nabla\times \nabla\times{\bf
\mathcal{G}}(x,x^{\prime})-k^2{\bf \mathcal{G}}(x,x^{\prime})={\bf
\mathcal{I}}_3\delta(x-x^{\prime}),\;\;& \mbox{  in }\R^3,  \\
\lim_{|x|\to \infty}|x|\big[ \nabla\times {\bf
\mathcal{G}}(x,x^{\prime})-ik\frac{x}{|x|}\times {\bf
\mathcal{G}}(x,x^{\prime}) \big]=0.&
\end{array}
\right .
$$
where ${\bf \mathcal{I}}_3$ is the $3\times 3$ identity matrix. In
the above notation the curl operator, $\nabla\times$, acts on
matrices column by column and $\nabla^{\prime}$ denotes the derivation with respect to
the second variable $x^{\prime}$.\\
The proof of Lemma \ref{lip} follows immediately if we multiply
the first equation in~(\ref{maxwell2}) by ${\bf
\mathcal{G}}(x,x^{\prime})\cdot {\bf v }$ $({\bf v }\in \R^3)$ and
if we integrate by parts over the domain $\Omega$ which contains
the set of inclusions $\{z_j + \alpha B_j\}_{j=1}^m$.
\section{Asymptotic behavior} As we said in the introduction our
analysis will be dependent on polarization tensors
$M\in\R^{3\times 3}$. We remember that this tensors are defined by
\begin{equation}\label{tensor1}
  M(q_j/q_0;B_j):=\ds\int_{B_j}\nabla v^{q}(x)dx,
\end{equation}
where $\{q_j\}$ (resp. $q_0$) denote either the set
$\{\varepsilon_j\}$ or $\{\mu_j\}$ for $1\leq j\leq m$ (resp.
denote either $\varepsilon_0$ or $\mu_0$) and where the functions
$v^{q}$, which dependent on the contrast $q_j/q_0$, are solutions
of the following problem, \begin{equation} \left \{
\begin{array}{*{2}{l}}
 \nabla\cdot q(x)\nabla v^{q}(x)=0,\;\;& \mbox{  in }\R^3,  \\
 v^{q}(x)-x\to 0 & \mbox{ as  }x\to\infty.
\end{array}\label{tensor2}
\right .
\end{equation}

The function $q$ is given by
$$
q(x)=\left \{ \begin{array}{*{2}{l}}
 q_0,\;\;& x\in\R^{3}\backslash \overline{B_j},  \\
 q_j,\;\;& x\in B_j.
\end{array}
\right .
$$
The following holds.
\begin{prop}\label{asym1}
Suppose that (\ref{f1}) and (\ref{tensor1}) are satisfied and let
$({\bf E}_{\alpha},
 {\bf H}_{\alpha})$ be the solution
 of the problem (\ref{maxwell1}), then we have
\begin{enumerate}
\item [(i)] The electric field ${\bf E}_{\alpha}$ satisfy the following uniform
expansion
 \begin{equation}\label{eqasym1}
 {\bf E}_{\alpha}(x)={\bf
E}_{0}(x)+\alpha^3\ds\sum_{j=1}^{m}\big[-i\omega(\mu_{j}-\mu_0)
\nabla^{\prime}\times{\bf \mathcal{G}}(x,z_j)\cdot{\bf
M}(\mu_{j}/\mu_0;B_j){\bf H}_{0}(z_j)
\end{equation}
$$
+\omega^2\mu_0(\varepsilon_{j}-\varepsilon_0){\bf
\mathcal{G}}(x,z_j)\cdot{\bf
M}(\varepsilon_{j}/\varepsilon_0;B_j){\bf
E}_{0}(z_j)\big]+O(\alpha^4).
 $$
\item [(ii)] The magnetic field ${\bf H}_{\alpha}$ satisfy the following uniform
expansion
\begin{equation}\label{eqasym2}
 {\bf H}_{\alpha}(x)={\bf
H}_{0}(x)+\alpha^3\ds\sum_{j=1}^{m}\big[i\omega(\varepsilon_{j}
-\varepsilon_0)\nabla^{\prime}\times{\bf
\mathcal{G}}(x,z_j)\cdot{\bf
M}(\varepsilon_{j}/\varepsilon_0;B_j){\bf E}_{0}(z_j)
\end{equation}
$$
 -\omega^2\varepsilon_0(\mu_{j}-\mu_0){\bf
\mathcal{G}}(x,z_j)\cdot{\bf M}(\mu_{j}/\mu_0;B_j){\bf
H}_{0}(z_j)\big] +O(\alpha^4).
 $$
\end{enumerate}
\end{prop}

To evaluate the influence of the presence of these inhomogeneities
with small diameters on the evolution of the electromagnetic
energy defined by~(\ref{energy-def}), we shall introduce its
density per unit of volume which denoted by $\aleph_\alpha$.
According to Poynting's theorem, energy density $\aleph_\alpha$ is
given by:
\begin{equation}\label{density1}
-~\ds\frac{\partial}{\partial t}\aleph_{\alpha}=\nabla\cdot{\bf
\Pi}_{\alpha}+{\bf J}_{s}\cdot{\bf E}_{\alpha},
\end{equation}
where ${\bf \Pi}_{\alpha}$ is the Poynting vector,
\begin{equation}\label{poyting1}
{\bf \Pi}_{\alpha}=\ds\frac{{\bf E}_{\alpha}\times {\bf
H}_{\alpha}}{\mu_0}.
\end{equation}

The following theorem is concerned with an important result which
justifies the behavior of the electromagnetic energy when a finite
number of inhomogeneities are introduced in the entire space.
\begin{thm}\label{density2}
Suppose that (\ref{f1}) and (\ref{density1}) are satisfied and
suppose that the inclusion $B_j$ is a ball for each $j\in
\{1,\cdots,m\}$. Then, the following uniform expansion holds \bea
\ds\frac{\partial}{\partial
t}(\aleph_{0}(x,t)-~\aleph_{\alpha}(x,t))&=&
\alpha^3\ds\sum_{j=1}^{m}3|B_j|\Big(\frac{\varepsilon_{j}-\varepsilon_0}
{\varepsilon_{j}+2\varepsilon_0}\Big[k^2{\bf J}_{s}\cdot({\bf
\mathcal{G}}(x,z_j)\cdot{\bf E}_{0}(z_j))\\ &&
+~1/\mu_0\nabla\cdot[i\omega\varepsilon_0{\bf
E}_{0}\times(\nabla^{\prime}\times{\bf
\mathcal{G}}(x,z_j)\cdot{\bf E}_{0}(z_j))\\ && +~k^2({\bf
\mathcal{G}}(x,z_j)\cdot{\bf E}_{0}(z_j))\times{\bf
H}_{0}(z_j)]\Big ]\\ && -~\frac{\mu_{j}-\mu_0}
{\mu_{j}+2\mu_0}\Big[i\omega\mu_0{\bf
J}_{s}\cdot(\nabla^{\prime}\times{\bf \mathcal{G}}(x,z_j)\cdot{\bf
H}_{0}(z_j))\\
&& +~1/\mu_0\nabla\cdot[i\omega\mu_0(\nabla^{\prime}\times{\bf
\mathcal{G}}(x,z_j)\cdot{\bf H}_{0}(z_j))\\ && +~k^2{\bf
E}_{0}\times{\bf \mathcal{G}}(x,z_j)\cdot{\bf
H}_{0}(z_j)]\Big]\Big) +O(\alpha^5). \eea
\end{thm}
\proof\\
Recall the following identity: $\nabla\cdot(A\times
B)=B\cdot\nabla\times A-A\cdot\nabla\times B.$ Relation
(\ref{poyting1}) immediately gives
\begin{equation}\label{poyting2} \nabla\cdot({\bf
\Pi}_{\alpha})=\ds\frac{1}{\mu_0}\big[{\bf
H}_{\alpha}\cdot\nabla\times{\bf E}_{\alpha} - {\bf
E}_{\alpha}\cdot\nabla\times{\bf H}_{\alpha}\big].
\end{equation}

Next, under assumption that inclusion $B_j$ is a ball for all
$j\in \{1,\cdots,m\}$, it was proved in~\cite{CMV} that
polarization tensors (\ref{tensor1}) are simplified
\begin{equation}\label{tensor3}
 M(\mu_{j}/\mu_0;B_j)= \frac{3\mu_0} {\mu_{j}+2\mu_0}|B_j|\mathcal{I}_3,\end{equation}

and

\begin{equation}\label{tensor4}M(\varepsilon_{j}/\varepsilon_0;B_j)= \frac{3\varepsilon_0}
{\varepsilon_{j}+2\varepsilon_0}|B_j|\mathcal{I}_3.\end{equation}

In other words, the inclusions $B_j$ are symmetric about their
centers (balls). Then, according to~\cite{ammari-darko} the
correction of order four in relation (\ref{eqasym1}) is zero and
therefore the remainder is in fact $O(\alpha^5)$. Using this
result and inserting relations (\ref{tensor3}) and (\ref{tensor4})
into relation (\ref{eqasym1}), the following expansions
immediately holds:
\begin{equation}\label{eqasym01}
 {\bf E}_{\alpha}(x)={\bf
E}_{0}(x)+\alpha^3\ds\sum_{j=1}^{m}\big[-i\omega\mu_0\frac{3(\mu_{j}-\mu_0)}
{\mu_{j}+2\mu_0)}|B_j|\nabla^{\prime}\times{\bf
\mathcal{G}}(x,z_j)\cdot{\bf H}_{0}(z_j)
\end{equation}
$$
+k^2\frac{3(\varepsilon_{j}-\varepsilon_0)}
{\varepsilon_{j}+2\varepsilon_0)}|B_j|{\bf
\mathcal{G}}(x,z_j)\cdot{\bf E}_{0}(z_j)\big] +O(\alpha^5),
 $$
In similar fashion we can get an expansion for the magnetic field:
\begin{equation}\label{eqasym02}
 {\bf H}_{\alpha}(x)={\bf
H}_{0}(x)+\alpha^3\ds\sum_{j=1}^{m}\big[i\omega\varepsilon_0\frac{3(\varepsilon_{j}
-\varepsilon_0)}
{\varepsilon_{j}+2\varepsilon_0)}|B_j|\nabla^{\prime}\times{\bf
\mathcal{G}}(x,z_j)\cdot{\bf E}_{0}(z_j)
\end{equation}
$$
 -
k^2\frac{3(\mu_{j}-\mu_0)} {\mu_{j}+2\mu_0)}|B_j|{\bf
\mathcal{G}}(x,z_j)\cdot{\bf H}_{0}(z_j)\big] +O(\alpha^5).
 $$
Using relations~(\ref{eqasym01}) and~(\ref{eqasym02}), the
following holds \bea {\bf H}_{\alpha}\cdot\nabla\times{\bf
E}_{\alpha}&=&{\bf H}_{0}\cdot\nabla\times{\bf
E}_{0}+\alpha^3\big\{\ds\sum_{j=1}^{m}c_{1}{\bf
H}_{0}\cdot\nabla\times(\nabla^{\prime}\times{\bf
\mathcal{G}}(x,z_j){\bf H}_{0})\\ && +~c_{2}{\bf
H}_{0}\cdot\nabla\times({\bf \mathcal{G}}(x,z_j){\bf E}_{0}) +
\ds\sum_{j=1}^{m}c_{1}^{\prime}(\nabla^{\prime}\times{\bf
\mathcal{G}}(x,z_j){\bf E}_{0})\cdot(\nabla\times{\bf E}_{0})\\
&& +~c_{2}^{\prime}({\bf \mathcal{G}}(x,z_j){\bf
H}_{0})\cdot(\nabla\times{\bf E}_{0}) \big\}+O(\alpha^5), \eea
where the constants $c_{1},~c_{2},~c_{1}^{\prime}$ and
$c_{2}^{\prime}$ are given by
$$
\left \{ \begin{array}{*{2}{l}}
 \ds\frac{c_{1}}{i\omega\mu_0}=\ds\frac{c_{2}^{\prime}}{k^2}=
    -3\ds\frac{\mu_{j}-\mu_0} {\mu_{j}+2\mu_0}|B_j|,&  \\
  \ds\frac{c_{2}}{k^2}=\ds\frac{c_{1}^{\prime}}{i\omega\varepsilon_0}=
    3\ds\frac{\varepsilon_{j}-\varepsilon_0} {\varepsilon_{j}
    +2\varepsilon_0}|B_j|.&
\end{array}
\right .
$$

In the same manner we find the relation for the term ${\bf
E}_{\alpha}\cdot\nabla\times{\bf H}_{\alpha}$; therefore
relation~(\ref{poyting2}) becomes \bea
 \nabla\cdot({\bf
\Pi}_{\alpha})&=&\nabla\cdot({\bf
\Pi}_{0})+\alpha^3\ds\sum_{j=1}^{m}\ds\frac{1}{\mu_0}
\Big[c_{1}\nabla\cdot\big((\nabla^{\prime}
\times{\bf \mathcal{G}}(x,z_j){\bf H}_{0})\times{\bf H}_{0}\big)\\
&& +~c_{2}\nabla\cdot\big({\bf \mathcal{G}}(x,z_j){\bf
E}_{0}\times{\bf H}_{0}\big) + c_{1}^{\prime}\nabla\cdot\big({\bf
E}_{0}\times(\nabla^{\prime}\times{\bf
\mathcal{G}}(x,z_j){\bf E}_{0})\big)\\
&& +~c_{2}^{\prime}\nabla\cdot\big({\bf E}_{0}\times{\bf
\mathcal{G}}(x,z_j){\bf H}_{0}\big) \Big]+O(\alpha^5). \eea

The proof is achieved by expanding the term ${\bf J}_{s}\cdot{\bf
E}_{\alpha}$ at order $5$ according to $\alpha$ in~(\ref{density1}). \square\\

Based on Theorem \ref{density2}, we can prove the following main
result.
\begin{thm}\label{estimation1}
Let $T>0$, then there exists some positive constant $C$ such that
the following estimate as $\alpha \rightarrow 0$ holds uniformly
in $t\in[0,T]$
$$
\ds|\mathcal{E}_{\alpha}(t)-\mathcal{E}_{0}(t)|\leq C\alpha^3,
$$
where the constant $C$ is independent of $\alpha$ and the set of
points $\{  z_j\}_{j=1}^m$ provided that assumption (\ref{f1})
holds, but this constant $C$ is dependent on $|B_j|$ and $T$.
\end{thm}


\vspace{1cm}

{\bf Christian DAVEAU}, -Adresse: {\it D\'epartement de
Math\'ematiques, Site Saint-Martin II,\\BP 222, \& Universit\'e de
Cergy-Pontoise, 95302 Cergy-Pontoise Cedex, France.}\\- Email:
christian.daveau@math.u-cergy.fr\\-Tel : (33) (0)1 34 25 66 72.
-Fax : (33) (0)1 34 25 66 45.\\

{\bf Abdessatar KHELIFI}, -Adresse: {\it D\'epartement de
Math\'ematiques, \& Universit\'e des Sciences de Carthage,
Bizerte, 7021, Tunisie.}\\ -Email: abdessatar.khelifi@fsb.rnu.tn\\
-Tel : (216) 97 53 17 13. -Fax : (216) 72 59 05 66.

\end{document}